\documentclass{article}
\usepackage{stackengine}
\usepackage[false]{anonymous-acm}
\usepackage{PRIMEarxiv}
\usepackage[utf8]{inputenc} % allow utf-8 input
\usepackage[T1]{fontenc}    % use 8-bit T1 fonts
\usepackage{hyperref}       % hyperlinks
\usepackage{url}            % simple URL typesetting
\usepackage{booktabs}       % professional-quality tables
\usepackage{amsfonts}       % blackboard math symbols
\usepackage{nicefrac}       % compact symbols for 1/2, etc.
\usepackage{microtype}      % microtypography
\usepackage{lipsum}
\usepackage{fancyhdr}       % header
\usepackage{graphicx}       % graphics
\graphicspath{{./}}     % organize your images and other figures under media/ folder
\usepackage{amsmath}
\usepackage{amssymb}
\usepackage{bm}
\usepackage{longtable}
\usepackage{circuitikz}
\usepackage{indentfirst}
\usepackage{cleveref}
\usepackage[inline,shortlabels]{enumitem}
\usepackage{siunitx}
\usepackage{caption}
\usepackage{subcaption}
\usepackage{multirow}
\usepackage{xcolor}
\usepackage[style=numeric-comp,sorting=none]{biblatex}
\bibliography{references}

\newlist{E}{enumerate}{1}
\setlist[E]{label=\textbf{E\arabic*:}}
\crefname{equation}{Eq.}{Eqs.}
\Crefname{equation}{Equation}{Equations}
\crefname{section}{Section}{Sections}
\Crefname{section}{Section}{Sections}
\crefname{figure}{Fig.}{Figs.}      
\Crefname{figure}{Figure}{Figures}  
\crefname{table}{Table}{Tables}
\Crefname{table}{Table}{Tables}
\newcommand\barbelow[1]{\stackunder[1.2pt]{$#1$}{\rule{1ex}{.075ex}}}
\newcommand{\myadd}[1]{\textcolor{black}{#1}}

\title{Control Policy Correction Framework for Reinforcement Learning-based Energy Arbitrage Strategies
\thanks{15th ACM International Conference on Future and Sustainable Energy Systems (ACM e-Energy 2024), Singapore. \\
https://doi.org/10.1145/3632775.3661948} 
}

\author{
  Seyed Soroush Karimi Madahi \\
  IDLab, Ghent University — imec \\
  \texttt{seyedsoroush.karimimadahi@ugent.be} \\
  \And
  Gargya Gokhale \\
  IDLab, Ghent University — imec \\
  \And
  Marie-Sophie Verwee \\
  IDLab, Ghent University — imec \\
  \And
  Bert Claessens \\
  BEEBOP \\
  \And
  Chris Develder \\
  IDLab, Ghent University — imec \\
}
\begin{document}
\maketitle

\begin{abstract}
  A continuous rise in the penetration of renewable energy sources, along with the use of the single imbalance pricing, provides a new opportunity for balance responsible parties to reduce their cost through energy arbitrage in the imbalance settlement mechanism. Model-free reinforcement learning (RL) methods are an appropriate choice for solving the energy arbitrage problem due to their outstanding performance in solving complex stochastic sequential problems. However, RL is rarely deployed in real-world applications since its learned policy does not necessarily guarantee safety during the execution phase. In this paper, we propose a new RL-based control framework for batteries to obtain a \myadd{safe} energy arbitrage strategy in the imbalance settlement mechanism. In our proposed control framework, the agent initially aims to optimize the arbitrage revenue. Subsequently, in the post-processing step, we correct (constrain) the learned policy following a knowledge distillation process \myadd{based on properties that follow human intuition.} Our post-processing step is a generic method and is not restricted to the energy arbitrage domain. We use the Belgian imbalance price of 2023 to evaluate the performance of our proposed framework. Furthermore, we deploy our proposed control framework on a real battery to show its capability in the real world.
\end{abstract}

\keywords{Battery energy storage systems, distributional reinforcement learning, energy arbitrage, interpretable reinforcement learning, knowledge distillation, safe reinforcement learning}

%===================================
\section{Introduction}
\label{sec:Introduction}
%===================================
Countries are making progress in transitioning toward a decarbonized electricity grid by adopting a larger amount of renewable energy sources (RES). However, the rise in shares of RES leads to an increasing mismatch between generation and consumption, given the dependence of RES generation on weather conditions. This mismatch poses challenges to transmission system operators (TSOs) in maintaining the balance of the grid. TSOs rely on balance responsible parties (BRPs) to assist in keeping supply and demand balance, by penalizing unbalanced BRPs in a periodic (typically 15\,min based) imbalance settlement scheme~\cite{baetens2020imbalance}. Europe's electricity balancing guideline (EBGL) stipulates calculating the imbalance cost based on a single imbalance price, implying that both negative and positive imbalances are penalized equally~\cite{ENTSO}. In this pricing method, BRPs can reduce their cost while assisting TSOs in maintaining grid balance by deviating from their day-ahead nomination. The given imbalance pricing scheme, and the increased need for balancing because of higher RES penetration presents a new opportunity for BRPs to reduce their cost through energy arbitrage in imbalance settlement.

The energy arbitrage problem is a \myadd{sequential} complex one, given the highly uncertain imbalance prices and the nearly real-time decision-making that is required. Most previous research is based on model-based optimization methods to obtain energy arbitrage strategies~\cite{Bottieau2020,Smets2023, Lago2021controller}. These methods formulate the energy arbitrage problem as a nonlinear programming problem which is typically non-convex, meaning that its optimal solution cannot be directly found. For this reason, linearization techniques (such as piecewise linear approximation) are used to transform the nonlinear problem into a linear or mixed-integer convex problem. However, using these linearization techniques might lead to an imprecise approximation or an intractable optimization problem. Stochastic optimization and robust optimization are the most popular model-based optimization methods. However, stochastic optimization has a high computational burden due to numerous scenarios, while a robust optimization solution tends to be extremely cautious~\cite{zhao2023day}. Reinforcement learning (RL) methods can deal with such model-based related problems: model-free RL methods do not require prior knowledge or a detailed model of the environment. The agent interacts with the environment to capture stochasticity in the environment and learn a (near-)optimal strategy. Also, RL does not have any specific hypothesis concerning the reward function adapting to any non-linear rewards. Furthermore, RL agents directly learn a control policy, without the need for repeatedly solving an optimization~(as seen in \myadd{model-based} methods), making them suitable for real-time control. RL methods have achieved state-of-the-art performance on many energy applications such as control of building systems~\cite{zhang2021joint}, electricity market modeling~\cite{harder2023modeling}, and voltage control~\cite{feng2023stability}.

In spite of the impressive performance of RL in simulations, it is challenging to deploy RL in real-world applications. Indeed, vanilla RL methods cannot guarantee the safety and correctness of the learned policy for unseen states. Safe RL aims to address this by maximizing the cumulative reward while satisfying safety constraints~\cite{achiam2017constrained}. Most safe RL methods try to constrain the policy during the learning process~\cite{chen2021enforcing,yang2021wcsac,alshiekh2018safe}. While effective, a major drawback of such methods is that, since the agent is trained to satisfy some predefined constraints, the learned constrained policy cannot be scaled to other similar settings: for redeploying the learned policy in other similar settings, the agent needs to be retrained (even if there is a slight change in the constraints). However, retraining RL agents is usually time-consuming and computationally expensive.

To reduce computation at inference time, knowledge distillation in neural networks was first introduced in~\cite{hinton2015distilling}. The key idea of knowledge distillation is that a larger teacher model transfers its knowledge to a smaller student model to achieve competitive performance and faster inference. Knowledge distillation is commonly used to address the huge computational burden and memory requirement of large models. It is shown that knowledge distillation improves model generalization as the student model is trained using soft targets instead of hard targets~\cite{tang2020understanding}. The teacher model tries during its training phase to assign the highest probability to the correct class (or best action) and small probabilities to incorrect classes (or other actions). These relative probabilities of incorrect classes provide more information about how the teacher model tends to generalize. Therefore, training the student model with soft targets transfers this generalization ability of the teacher model efficiently to the student model.

In this paper, we introduce the extension of the standard knowledge distillation process by adding an optimization layer to the student model to correct and constrain the RL policy. Building upon this idea, we propose a new RL-based control framework for batteries to obtain a \myadd{safe} energy arbitrage strategy in the imbalance settlement mechanism. In the proposed control framework, the agent is initially trained to maximize the arbitrage profit (\cref{sec:Problem Formulation}). Afterwards, in the post-processing step, the aim is to transfer knowledge from the trained agent to the student agent while correcting the policy of the student agent (\cref{sec:Policy Correction}). The policy correction of the student agent is based on human intuitive constraints\myadd{,} to make the final policy \myadd{rational} from a human perspective. The main advantage of the proposed framework over previous online safe RL methods lies in its greater flexibility for reusing pretrained RL agents. By applying the post-processing step, BRPs can avoid retraining from scratch and effortlessly reuse pretrained RL agents in accordance with their own defined constraints and preferences. 

We employ distributional deep Q learning (DDQN), a state-of-the-art RL method. The main advantage of distributional RL over standard RL is its ability to estimate the complete probability distribution of returns instead of relying on a single value expectation, resulting in superior performance. However, the proposed policy correction process and control framework can be applied to all RL methods. The performance of the proposed control framework is validated using the Belgian imbalance price of 2023. Moreover, a real-time experimental study using a real battery is conducted to better demonstrate the capability of the proposed control framework in the real world (\cref{sec:Results,sec:Conclusion}). Our main contributions in this paper are\myadd{:}
\begin{itemize}
    \item We introduce a new policy correction step that can be applied to any RL method to ensure the correctness and safety of the final policy; 
    \item We propose a distributional RL-based control framework for a battery to obtain a \myadd{safe} energy arbitrage strategy in the imbalance settlement mechanism \myadd{based on properties that adhere to human intuition};
    \item We implement the proposed control framework on a real battery to evaluate its performance in the real world.
\end{itemize}

%===================================
\section{Background and Related Work}
\label{sec:Background}
%===================================
\textbf{Imbalance Settlement.} BRPs are responsible for consistently balancing their individual consumption and generation. However, a deviation from their day-ahead nomination is inevitable because of uncertainties in RES. To correct the system imbalance, a TSO activates reserve capacities offered in the balancing market, and BRPs incur charges from the TSO for their imbalance at the end of the imbalance settlement period (15 mins in most European countries) ~\cite{Lago2021market}. This mechanism is called imbalance settlement. The imbalance price is determined based on the reserve volume and direction activated by the TSO~\cite{vatandoust2023integrated}. Two main imbalance pricing methodologies are used in different countries: (1) \textit{dual pricing}, where the price is different for positive and negative imbalances, and (2) \textit{single pricing}, where the price is the same for both imbalance directions and is determined by the total system imbalance. As mentioned earlier, the objective of ENTSO-E is to standardize the imbalance settlement mechanism in Europe by adopting the single pricing method for calculating the imbalance price for each 15-minute imbalance settlement period. Hence, we focus on the single pricing methodology. The Belgian imbalance settlement mechanism was chosen as a case study for our research~\cite{baetens2020imbalance}.

Energy arbitrage refers to a technique to gain financial profits by buying energy at lower prices and selling it at higher prices. Due to the high volatility in imbalance prices and the need for near real-time \myadd{sequential} decision-making, energy arbitrage is challenging. For this reason, together with the recent change in the imbalance pricing methodology, only few research studies have been conducted on the arbitrage in the imbalance settlement mechanism~\cite{Bottieau2020,Smets2023,madahi2023distributional,Lago2021controller}. A new tailored encoder-decoder architecture was implemented in~\cite{Bottieau2020} to generate improved probabilistic predictions of the future system imbalance. Afterward, a bi-level robust optimization problem was solved to optimize the profit of a BESS in the imbalance settlement. A novel risk-aware stochastic model predictive control (MPC) approach was introduced in~\cite{Smets2023} to maximize the revenue of BESS in the imbalance settlement mechanism while considering battery degradation costs.

%The authors in~\cite{madahi2023distributional} propose a battery control framework based on distributional RL for a risk-sensitive energy arbitrage in the imbalance settlement mechanism, taking into account a cycle constraint. 

Most of the mentioned studies have solved the arbitrage problem by applying model-based optimization methods. However, the main drawback of these methods is that they need linearization techniques to estimate the nonlinear problem as a linear (or mixed-integer) convex problem that can lead to an inaccurate approximation. To address such problems of model-based optimization methods, we deploy model-free RL methods. %\myadd{The main difference between our work and~\cite{madahi2023distributional} is the post-processing policy correction step. Our proposed solution focuses on safety and attempts to correct actions during the knowledge distillation process through adopting implicit optimization layer. Nevertheless, the primary focus of~\cite{madahi2023distributional} was to achieve state-of-the-art performance in the arbitrage problem and introduce a data-driven risk-sensitive framework. Furthermore, in this paper, we evaluate the performance of the proposed control framework on a real battery.
\myadd{Several other research works %have been conducted to perform 
already studied energy arbitrage using model-free RL. For example, \cite{madahi2023distributional}~proposes a battery control framework based on distributional RL for a risk-sensitive energy arbitrage in the imbalance settlement mechanism, taking into account a cycle constraint.
In~\cite{cao2020deep}, an RL-based method was proposed to optimize battery energy arbitrage in the day-ahead market, taking into account an accurate battery degradation model. The authors in~\cite{li2022learn} present an RL-based battery bidding strategy in the real-time and frequency control ancillary services markets, using a transformer-based temporal feature extractor. In~\cite{qi2019energyboost}, learning-based control algorithms were proposed to obtain an optimal policy for home batteries. Further, \cite{xu2024optimal} used a deep-RL approach to solve the electricity arbitrage problem in the day-ahead market.}
\par \myadd{Our current work is complementary to all of the above works on RL for energy arbitrage\textanon{, and particularly to our earlier research in distributional RL~\cite{madahi2023distributional}. The latter's main objective was to establish which distributional RL algorithm (distributional SAC vs.\ DQN) performed best, for a battery controller with cycle constraints in imbalance settlement. Yet,}{. Indeed,}
our current work specifically looks at the problem that learned RL policies may exhibit erratic behavior, in that they take surprising (if not non-optimal) actions in certain regions of the system state space, especially those that have been only infrequently visited during training. We specifically propose a post-processing step to correct those policies, adopting a distillation setup.}

%Such behavior can be exhibited by distributional RL solutions such as those explored in~\cite{madahi2023distributional}.
%Previous works did not specifically consider that problem: e.g., \cite{madahi2023distributional}'s primary goal was to establish which distributional RL algorithms performed best in energy arbitrage, and particularly for a battery controller taking into account cycle constraints.

\textbf{Safe RL.} In RL literature, the concept of safety is used in opposition to risk, and it is not solely confined to physical damage~\cite{garcia2015comprehensive}. In stochastic environments, the learned optimal policy may result in poor performance as the learned policy is not necessarily robust against the rare occurrence of large negative returns. The risk stems from uncertainties in the environment. Some safe RL works focused on \textit{domain knowledge} of the problem to guarantee safety, such as safe exploration~\cite{dalal2018safe}, designing a safety shield~\cite{alshiekh2018safe}, and human interventions~\cite{saunders2017trial}. In these works, the safety model is assumed to be known priori. On the other hand, there is another class of safe RL work that focus on \textit{constrained} optimization. In these studies, the safe RL problems are commonly formulated as a constrained Markov decision process (CMDP) (e.g.,~\cite{achiam2017constrained}). Usually, a Lagrange-multiplier method is applied to transform the constrained problem into a non-constrained one (e.g.,~\cite{yang2021wcsac}). In~\cite{chen2021enforcing}, convex constraints are enforced in learned policies by incorporating a differentiable projection layer within a neural network-based policy. The authors in~\cite{ruelens2016residential} repair the RL policy using expert knowledge by solving a convex optimization problem. They define the optimization problem using a fuzzy model with triangular membership functions to approximate the policy. \myadd{In~\cite{rahman2022efficient}, an implicit optimization layer was employed for projecting the actions taken by an RL-based controller to ensure satisfaction of electric vehicle constraints.}

The main disadvantage of the previous safe RL works is that they generally constrain the policy during the \textit{training phase}. Thus, it is not straightforward to (re)deploy the learned constrained policy in other similar settings with slight changes in constraints. A new agent needs to be trained for each new setting, even if changes in constraints are minor (e.g., boundary values). In some domains, it is easy to distinguish sub-optimal actions from catastrophic actions -- for instance, an autonomous vehicle driving too slowly is sub-optimal, while the car driving into a group of pedestrians is clearly catastrophic. However, in the energy arbitrage domain, the distinction between sub-optimal and catastrophic actions is challenging, since safety and risk preferences may significantly differ among BRPs. This problem becomes more pronounced when a safe agent is trained for energy arbitrage in a highly volatile market, such as the imbalance settlement mechanism. We therefore propose a \textit{new post-processing} step to constrain (correct) a pretrained unconstrained policy. The main advantage of our framework is that it avoids retraining the agent from scratch, a process that is typically time-consuming and computationally expensive. Instead, our proposed framework constrains pretrained agents according to the risk preference of BRPs in a post-processing step. This provides an opportunity to reuse a pretrained agent in similar settings with different constraints. It is worth noting that our proposed post-processing step can be applied to any RL method.

\textbf{Knowledge distillation.} It is usually preferred to train cumbersome deep neural networks with a strong regularizer such as dropout because overparameterization increases the generalization performance~\cite{gou2021knowledge}. However, their computational complexity and slow inference limit their usage in many applications. Knowledge distillation is a technique used to efficiently compress the capabilities of a larger teacher model into a smaller student model. Different forms of knowledge can be transferred from the teacher model to the student model, e.g., logits~\cite{hinton2015distilling}, feature maps~\cite{passban2021alp}, and relations between pairs of feature maps~\cite{yim2017gift}. Apart from compaction, knowledge distillation is also used to improve generalization~\cite{tang2020understanding}, reproducibility~\cite{anil2018large}, data augmentation~\cite{lee2020self}, and defend from adversarial attacks~\cite{papernot2016distillation}. The teacher model based on its confidence in the ground-truth class rescales gradients of the student model. Also, the teacher model’s probability mass on incorrect classes reflects class relationships, offering more guidance to the student model~\cite{tang2020understanding}. These effects contribute to improving the generalization of the student model. In this paper, we correct (constrain) the policy when the logits knowledge is transferred from the pretrained unconstrained neural network-based policy to the student model.

%===================================
\section{Problem Formulation}
\label{sec:Problem Formulation}
%===================================
In this section, the energy arbitrage problem in the imbalance settlement mechanism is formulated as a Markov decision process (MDP) (\cref{sec:MDP}) and the RL method used for solving the problem is explained in detail (\cref{sec:DDQN}).

%--------------------------------
\subsection{MDP Formulation}
\label{sec:MDP}
%--------------------------------
An MDP presents stochastic sequential decision-making problems as a mathematical framework. The MDP problem is defined by a tuple $(\mathcal{S},\mathcal{A},\mathcal{R},\mathcal{P},\gamma)$, where $\mathcal{S}$ represents the state space, $\mathcal{A}$ denotes the (discrete) action space, $\mathcal{R}: \mathcal{S}\times\mathcal{A}\rightarrow\mathbb{R}$ is the instantaneous reward function, $\mathcal{P}: \mathcal{S}\times\mathcal{S}\times\mathcal{A}\rightarrow[0,1]$ represents the unknown state transition probability distribution, and a discount factor $\gamma \in (0,1]$~\cite{Sutton2018}. At each time step $t$, the environment state $s_t \in \mathcal{S}$ is the observation of the agent. After taking action $a_t \in \mathcal{A}$, the environment provides a reward value $\mathcal{R}(s_t,a_t)$ for the agent. The state transition probability distribution $\mathcal{P}(s_\text{t+1}|s_t,a_t)$ determines the probability of moving to a new state $s_\text{t+1} \in \mathcal{S}$. \myadd{In our energy arbitrage problem, an instantaneous action taken by the agent affects the overall expected profit of the agent. If the agent prioritizes immediate rewards and discharges the battery completely, it can miss high price periods and make less overall profit.} The agent in the energy arbitrage problem makes a decision (i.e., action) at each time step regarding the charging/ discharging of battery. Electricity markets and the grid are considered the environment the agent interacts with. We define the MDP formulation for the energy arbitrage problem in the imbalance settlement mechanism as:
\begin{enumerate}[(i)]
\item \emph{State}: The state at each time step (which is considered to be 2 minutes) is given by
\begin{equation}
    s_t=(T_\textrm{qh},\textit{qh},\textit{mo},\textit{SOC}_t,\hat{\pi}^\textrm{imb}_t)
    \label{1}
\end{equation}
where $T_\textrm{qh} \in [0,14] $ is the minute of the quarter hour, $\textit{qh} \in [0,95]$ denotes the quarter hour of the day, $\textit{mo}$ represents the month of the year, $\textit{SOC}_t$ is the state of charge (SoC) of battery at time $t$. Finally, $\hat{\pi}^\textrm{imb}_t$ is the \myadd{\emph{indicative}} imbalance price of the current quarter hour $\textit{qh}$. \myadd{Indeed,} the %real
\myadd{\emph{actual}} imbalance price of each quarter hour is only known at the end of the quarter hour. For this reason, our agent can only observe \myadd{an indicative imbalance price} for the current quarter hour. \myadd{Due to uncertainty in the imbalance price of the current quarter hour, our defined arbitrage problem is stochastic.}

\item \emph{Action}: Our action space is discrete, consisting of three possible actions, defined as follows:
\begin{equation}
    a_t \in \mathcal{A}, \quad \mathcal{A}=\{-P_\text{max}, 0, P_\text{max}\}
    \label{2}
\end{equation}
where  $P_\text{max}$ is the maximum (dis-)charging power of the battery. A positive action $a_t$ corresponds to charging the battery, while a negative action means discharging the battery at time $t$. \myadd{The emergence of bang-bang behavior was investigated in continuous control RL by~\cite{seyde2021bang}. They showed that RL methods with discrete action space can achieve competitive performance on standard continuous control benchmarks. For this reason, we use a discrete action space in this paper.}
\item \emph{Reward}: The agent aims to maximize the profit by purchasing energy at cheap imbalance prices and selling it at expensive imbalance prices. Therefore,the reward function is defined as the negative of the energy cost:
\begin{equation}
    r_t=-a_t\pi^\textrm{imb}_\textrm{qh},
    \label{3}
\end{equation}
where $\pi^\textrm{imb}_\textrm{qh}$ represents the real imbalance price of the quarter hour in which $t$ lies.
\item \emph{State transition function}: A state transition probability function $\mathcal{P}$ describes system dynamics. This probability function is unknown for the agent in our problem due to uncertainties in the imbalance price. More specifically, the probability distribution of $\hat{\pi}^\textrm{imb}_\text{t+1}$ given $\hat{\pi}^\textrm{imb}_t$ (P($\hat{\pi}^\textrm{imb}_\text{t+1}$|$\hat{\pi}^\textrm{imb}_t$)) is the only source of stochasticity in the agent state and it is independent from the taken action. Nevertheless, the state transition for $\textrm{SOC}_t$ is explicitly calculated as shown below, since it is influenced by $a_t$.
\begin{equation}
    \textrm{SOC}_\text{t+1} = \begin{cases}
        \textrm{SOC}^\text{temp}_\text{t+1} &: 0<\textrm{SOC}^\text{temp}_\text{t+1}<1 \\
        0 &: \textrm{SOC}^\text{temp}_\text{t+1}<0 \\
        1 &: \textrm{SOC}^\text{temp}_\text{t+1}>1
    \end{cases}
    \label{4}
\end{equation}
\begin{equation}
    \textrm{SOC}^\text{temp}_\text{t+1}=\textrm{SOC}_t+(\max(a_t,0) \eta_\text{cha}+\frac{\min(a_t,0)}{\eta_\text{dis}})\frac{\Delta t}{C_\text{BESS}}
    \label{5}
\end{equation}
where $C_\text{BESS}$ is the maximum capacity of the battery, and $\eta_\text{cha}$ and $\eta_\text{dis}$, are the charging and discharging efficiency ($\in [0,1]$) of the battery, respectively. The interaction between the agent and the environment helps the agent to estimate the transition probability distribution. Although the part of the state transition related to the SoC is deterministic, the agent is unaware of it and needs to learn it. 
\end{enumerate}

%--------------------------------
\subsection{Distributional Deep Q Learning}
\label{sec:DDQN}
%--------------------------------
We will solve the arbitrage problem, formulated as the MDP above, using a RL method. RL agents learn a policy to maximize the expected long-term reward. Classical tabular RL methods, e.g., Q-learning, cannot be applied to problems with high-dimensional or continuous state space due to the curse of dimensionality. Moreover, for replacing the tabular Q values with a function approximator, features need to be manually extracted~\cite{cao2020reinforcement}. To deal with these limitations, the deep Q learning (DQN) method suggests an idea of using a deep neural network as a function approximator to estimate the Q function. To learn the Q function $Q_\theta(s_t,a_t)$, the following loss function is minimized:
\begin{equation}
    L_Q(\theta)=\mathbb{E}_{(s_t,a_t,r_t,s_\text{t+1}) \sim \mathcal{D}} \left[(r_t+\gamma \max_a Q_\text{$\theta'$}(s_\text{t+1},a)-Q_\theta(s_t,a_t))^2 \right].
    \label{6}
\end{equation}
\begin{equation}
    \theta'=\tau \theta + (1-\tau) \theta'
    \label{7}
\end{equation}
The DQN method is stable in learning because of using the target Q function $Q_\text{$\theta'$}(s_t,a_t)$ for calculating next state-action values in \cref{6}, and training by a mini-batch sampled from an experience replay buffer $\mathcal{D}$~\cite{mnih2015human}. Furthermore, this method can avoid overfitting the policy because it is an off-policy method that can learn from historical data, not just current experiences~\cite{srivastava2014dropout}.

The authors in~\cite{bellemare2017distributional} introduced for the first time a distributional perspective on RL. These methods learn the probability distribution over returns instead of a single-value return. Distributional RL methods have various benefits, including the mitigation of Q-value overestimation~\cite{duan2021distributional}, facilitating learning risk-sensitive polices~\cite{theate2023risk}, and improving training stability~\cite{bellemare2017distributional}.

Extending beyond the basic DQN method, in DDQN, the probability distribution of returns ($\mathcal{Z}_\theta$) is learned using the distributional Bellman equation below~\cite{bellemare2017distributional}:
\begin{equation}
L_\mathcal{Z}(\theta)=\mathbb{E}_{(s_t,a_t) \sim \mathcal{D}}[D_\text{KL}(\mathcal{T} \mathcal{Z}_\text{$\theta'$}(s_t,a_t)||\mathcal{Z}_\theta(s_t,a_t))]
\label{8}
\end{equation}
\begin{equation}
\mathcal{T}Z(s_t,a_t)\overset{D}{=}r_t+\gamma \max_a \mathbb{E}_{Z \sim \mathcal{Z}_\text{$\theta'$}}[Z(s_\text{r+1},a)]
    \label{9}
\end{equation}
In \cref{8,9} $\mathcal{T}\mathcal{Z}_\theta$ is the probability distribution of $\mathcal{T}Z$, and $A\overset{D}{=}B$ indicates the equality of probability distributions for two random variables $A$ and $B$. $D_\text{KL}(.)$ denotes Kullback-Leibler (KL) divergence loss. we formulate the distribution over returns $Z$ as a categorical distribution:
\begin{equation}
Z(s_t,a_t)= \left\{ z_i \Big| z_i=V_\text{min}+\frac{V_\text{max}-V_\text{min}}{N-1}i, 0 \leq i < N\right \}
    \label{10}
\end{equation}
where $V_\text{min}$ and $V_\text{max}$ are the maximum and minimum values of random returns, respectively, and $N$ is the number of bins.

%===================================
\section{Policy Correction}
\label{sec:Policy Correction}
%===================================
As mentioned earlier, vanilla RL methods cannot guarantee good performance in improbable states due to several reasons. First, as these states occur rarely (or sometimes they do not exist in the training set), the agent cannot learn a good action for them and tries to take action based on the agent’s generalization ability. Second, the agent takes an action to maximize the \textit{expected} return, not based on the \textit{worst-case scenario} return, leading to taking catastrophic actions. In such a case, the agent might observe a few times that taking the specific action can cause a large negative return. However, since most of the time taking this action results in a high positive return, the agent still prefers this action over other possible actions. In this section, we propose a post-processing step to correct these catastrophic actions in our energy arbitrage problem. The purpose is to post-process the learned RL policy in a way that yields a correct and interpretable policy from a human perspective.

A human-intuitive energy arbitrage policy for our problem needs to possess three key properties: (1) charge the battery at very low prices; (2) discharge the battery at very high prices; (3) be monotonic with respect to price and SoC. The first two properties refer to the fact that the agent always must react properly to extremely rare prices regardless of the SoC level. The third property is the most crucial feature that makes the policy human-intuitive. For instance, we expect that when the agent decides to charge the battery at a specific price, it should also charge the battery at prices lower than this specific price, assuming other elements in the state remain constant. The following optimization problem ensures that the resulting neural network-based policy $\pi_\varphi$ has all the above-mentioned properties.
\begin{equation}
\begin{aligned}
    & \min_{\varphi} \quad{\myadd{\mathcal{L}_d}[\mu_\theta(s_t), \pi_\varphi(s_t)]} \;\myadd{, s_t \in \mathcal{S}} \\
    \textrm{s.t.} \quad \textrm{Property 1:} \quad & \textrm{argmax} \; \pi_\varphi(s_t) = P_\text{max} \,(\textrm{charging}) \;, \\ 
    & \textrm{if} \quad \hat{\pi}^\textrm{imb}_t \leq \ \barbelow{\pi}^\textrm{imb} \\
    \textrm{Property 2:} \quad & \textrm{argmax} \; \pi_\varphi(s_t) = -P_\text{max} (\textrm{discharging}) \;,\\
    & \textrm{if} \quad \hat{\pi}^\textrm{imb}_t \geq \ \overline{\pi}^\textrm{imb} \\
    \textrm{Property 3:} \quad & \textrm{argmax} \; \pi_\varphi(s_t) \geq \textrm{argmax} \; \pi_\varphi(s'_t) \;, \\
    & \textrm{if} \quad \textrm{SOC}_t \leq \textrm{SOC}'_t \; \textrm{and} \; \hat{\pi}^\textrm{imb}_t \leq \hat{\pi'}^\textrm{imb}_t
    \label{11}
\end{aligned}
\end{equation}
In \cref{11}, $\barbelow{\pi}^\textrm{imb}$ and $\overline{\pi}^\textrm{imb}$ are lower and upper bound thresholds for the imbalance price, respectively,  \myadd{$\mathcal{L}_d(.)$ is a loss function defined in~\cref{12}}, and $\mu_\theta(.)$ is the unconstrained neural network-based (pretrained) policy. Note that for the last property, other elements in $s_t$ and $s'_t$ are the same.

\begin{figure}[h]
    \centering
    \includegraphics[width=0.49\linewidth]{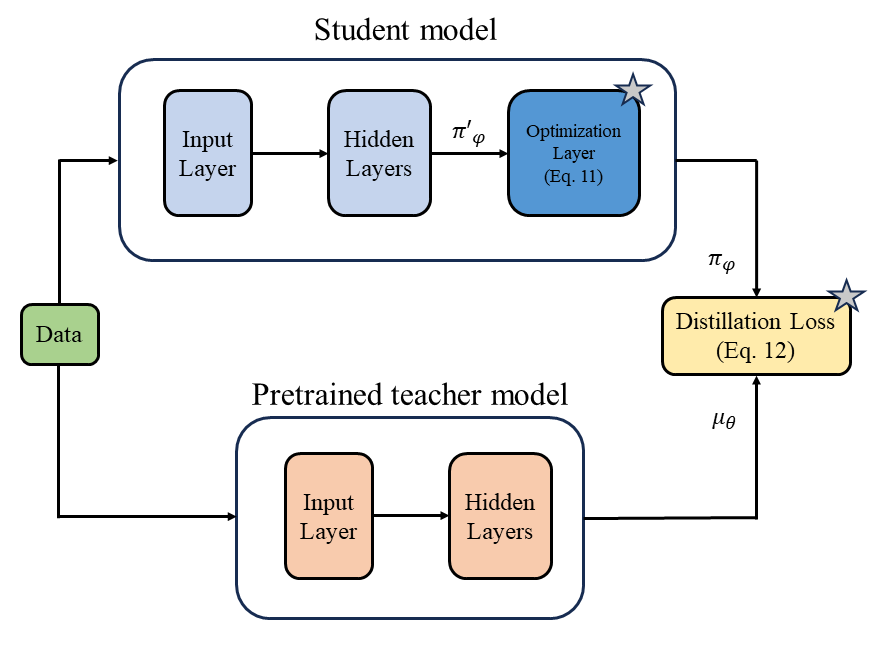}
    \caption{The proposed extension of knowledge distillation for policy correction. The starred blocks indicate our new addition to the standard distillation process.}
    \label{fig:KD}
\end{figure}

To solve the optimization problem in \cref{11}, we formulate it as a knowledge distillation process. \Cref{fig:KD} shows the proposed framework for knowledge distillation. During knowledge distillation, the teacher model $\mu_\theta$ is fixed and only the student model $\pi_\varphi$ is trained. The last layer of the student model is a differentiable optimization layer. The optimization layer enforces the three human-intuitive properties into the final policy. \myadd{As these three properties are formulated as a least squares problem with linear constraints in this paper, the resulting optimization problem is convex and can be solved using the optimization layer.} \myadd{We use the} \emph{cvxpylayers} package %is used
to implement this optimization layer~\cite{agrawal2019differentiable}. In the forward pass, the optimization layer solves the convex optimization problem, while the weights of the student model are trained end-to-end based on the gradient of the optimization solution during backpropagation. 

\begin{equation}
    \myadd{\mathcal{L}_d[\mu_\theta(s_t), \pi_\varphi(s_t)]} = \omega D_\text{KL}(\pi_\varphi || \mu_\theta) +  D_\text{KL}(\pi'_\varphi || \pi_\varphi)\\
    \label{12}
\end{equation}
\myadd{\Cref{12} formulates the loss function in \cref{11}} where $\pi'_\varphi$ is the output of the student model before the optimization layer and $\omega$ is a parameter to balance the two KL divergence losses. The first term in the loss function ensures that the student model closely mimics the behavior of the teacher model. On the other hand, the second term tries to keep the output of the student model before and after the optimization layer as close as possible. Although the optimization layer provides considerable flexibility in imposing convex constraints on the final policy, its computational burden to the student model makes the model inefficient, especially for real-world applications. To address this problem, the second term is added to the loss function. In that way, the student model is forced to train its weights to meet constraints in the absence of the optimization layer making the policy $\pi'_\varphi$ as close as possible to the constrained one $\pi_\varphi$. Consequently, during inference, the optimization layer can be ignored, and the output of the student model before optimization layer ($\pi'_\varphi$) is used for decision-making. 

\myadd{Our intuition for using knowledge distillation for the policy correction %can be highlighted from two perspectives.
has two parts.
First, knowledge distillation provides a framework for reusing pretrained RL agents. In this way, BRPs can avoid retraining from scratch and effortlessly reuse pretrained RL agents based on their own defined constraints and preferences. Second, the optimization layer is computationally expensive. In this arbitrage problem, the agent requires \num{50000} episodes to converge. Considering a minibatch size of \num{16384}, training the teacher model with the optimization layer for \num{50000} episodes would drastically increase the runtime to a point where the training loop may become intractable. On the other hand, in the knowledge distillation process, the student model only needs 600 epochs to be trained. Therefore, knowledge distillation can significantly reduce the computational time in our arbitrage problem.}

In the end, it is noteworthy to highlight that our proposed post-processing step is a generic framework and it is not limited to energy domain applications. Using our proposed framework, the policy of pretrained RL agents can be corrected by a set of defined convex constraints. Moreover, the optimization problem defined in \cref{11} is an example of constraints (rules) that can make the final energy arbitrage policy interpretable to humans. BRPs can have their individual preferences and define their own set of constraints. Also, these constraints depend on the inputs of the controller, and therefore can be adapted to those inputs.

%===================================
\section{Results}
\label{sec:Results}
%===================================
In this section, we evaluate the performance of our proposed control framework, as explained in \cref{sec:Problem Formulation,sec:Policy Correction}, through simulation and experimental results.

%--------------------------------
\subsection{Simulation Setup}
\label{sec:Simulation Setup}
%--------------------------------
We use the Belgian imbalance price of 2023 for evaluating our proposed framework.\footnote{https://opendata.elia.be/pages/home/} BRPs imbalances are settled at 15-minute-based prices, which are calculated at the end of each quarter-hour period. To provide more information to BRPs, the TSO (Elia) also publishes 1-minute-based \myadd{indicative} prices in real-time which are calculated based on the instantaneous system imbalance and prices of cumulative activated regulation volumes on a minute basis.\footnote{https://www.elia.be/-/media/project/elia/elia-site/grid-data/balancing/20190827\_end-user-documentation-elia1-minute-publications.pdf} We use these non-validated prices as a \myadd{indicator} of the real imbalance price for the related quarter hour period. \myadd{By getting closer to the end of the quarter hour, the indicative price becomes closer to the real imbalance price. Thus, the time resolution of decision-making needs to be sufficiently short to benefit from the most recent situation of the grid using the indicative price and to take action accordingly. The resolution for the RL agent is \myadd{hence} set to 2\,min.} The characteristics of the simulated battery are 4MW/ 8MWh with 90\% round-trip efficiency. To increase the lifetime of the battery, a minimum SoC is set at 10\%. For the sake of proof-of-concept study, we assume that the battery does not participate in the day-ahead \myadd{or other} markets. 

To train the initial unconstrained RL agent, the price dataset is split as follows: the first 20 days of each month as a training set, the 21st to the 25th of each month as a validation set, and the remaining days as a test set. The RL methods are trained with \num{50000} episodes, where each episode constitutes a single day. The discount factor $\gamma$, the soft update factor $\tau$, the experience replay buffer size, and the mini-batch size are set to \num{0.999}, 0.1, \num{1e6}, and \num{16384}, respectively. In both vanilla DQN and DDQN methods, the Q-value function and target Q function are modeled by a fully connected neural network that has two hidden layers with 256 and 128 neurons, respectively. The learning rate of the networks is \num{5e-4}. In the DDQN method, $V_\text{max}=-V_\text{min}=\num{1e5}$ and $N=51$. 

For the knowledge distillation process, a 2-layer fully connected neural network with hidden layer dimensions of 64 and 32 is used as the student model $\pi_\varphi$. The teacher model architecture is the same as the model trained in the previous step. The student model is trained for 600 epochs with the learning rate of \num{1e-3} using the Adam optimizer. \myadd{Tuning $\omega$ plays an important role in the training of the student model. Choosing large $\omega$ results in a student model that more frequently violates constraints. Conversely, a small $\omega$ leads to a student model that is less similar to the teacher model. To keep this balance, we used $\omega=\num{1e-4}$.} The values of $\barbelow{\pi}^\textrm{imb}$ and $\overline{\pi}^\textrm{imb}$ in \cref{11}, are set to $-$500 and 1500, respectively. The main decision boundaries for the decision-making are learned by the RL agent: $\barbelow{\pi}^\textrm{imb}$ and $\overline{\pi}^\textrm{imb}$ just define safety thresholds for the agent. For this reason, these safety thresholds must correspond to extremely rare prices to avoid impacting the main policy and primary decision boundaries. However, the definition of these safety thresholds can vary among BRPs due to their different risk preferences. The PyTorch package in Python is used to implement our proposed control framework.

\myadd{To benchmark the proposed control framework, a rule-based controller (RBC) is introduced as a baseline method. The RBC in this paper is a rudimentary threshold-based controller with two cutoff points based on the statistical analysis of imbalance prices. Although the RBC does not guarantee the optimal operation of the battery, it is widely used in the real-world due to its simplicity and ease of implementation~\cite{qi2019energyboost}. These thresholds classify imbalance prices into three categories: cheap, normal, and expensive. In the RBC, the battery is (dis)charged with the maximum power when the price is (above) below a (upper) lower bound. The RBC action is formulated as follows:
\begin{equation}
    a_t = \begin{cases}
        P_\text{max} &: \hat{\pi}^\textrm{imb}_t<\barbelow{\lambda} \\
        0 &: \barbelow{\lambda} \leq \hat{\pi}^\textrm{imb}_t \leq \overline{\lambda} \\
        -P_\text{max} &: \hat{\pi}^\textrm{imb}_t>\overline{\lambda}
    \end{cases}
    \label{13}
\end{equation}
where $\barbelow{\lambda}$ and $\overline{\lambda}$ represent lower and upper bounds, respectively. These thresholds are determined according to the distribution of the Belgian imbalance price in 2023: %. In this study,
we set the upper and lower bounds %are set
to the first and third quartiles of the 2023 prices, which are equal to 11~{\texteuro}/MWh and 179~{\texteuro}/MWh, respectively.}

%--------------------------------
\subsection{Simulation Results}
\label{sec:Simulation Results}
\begin{figure}[h]
    \centering
    \begin{subfigure}{0.49\textwidth}
        \includegraphics[scale=0.5]{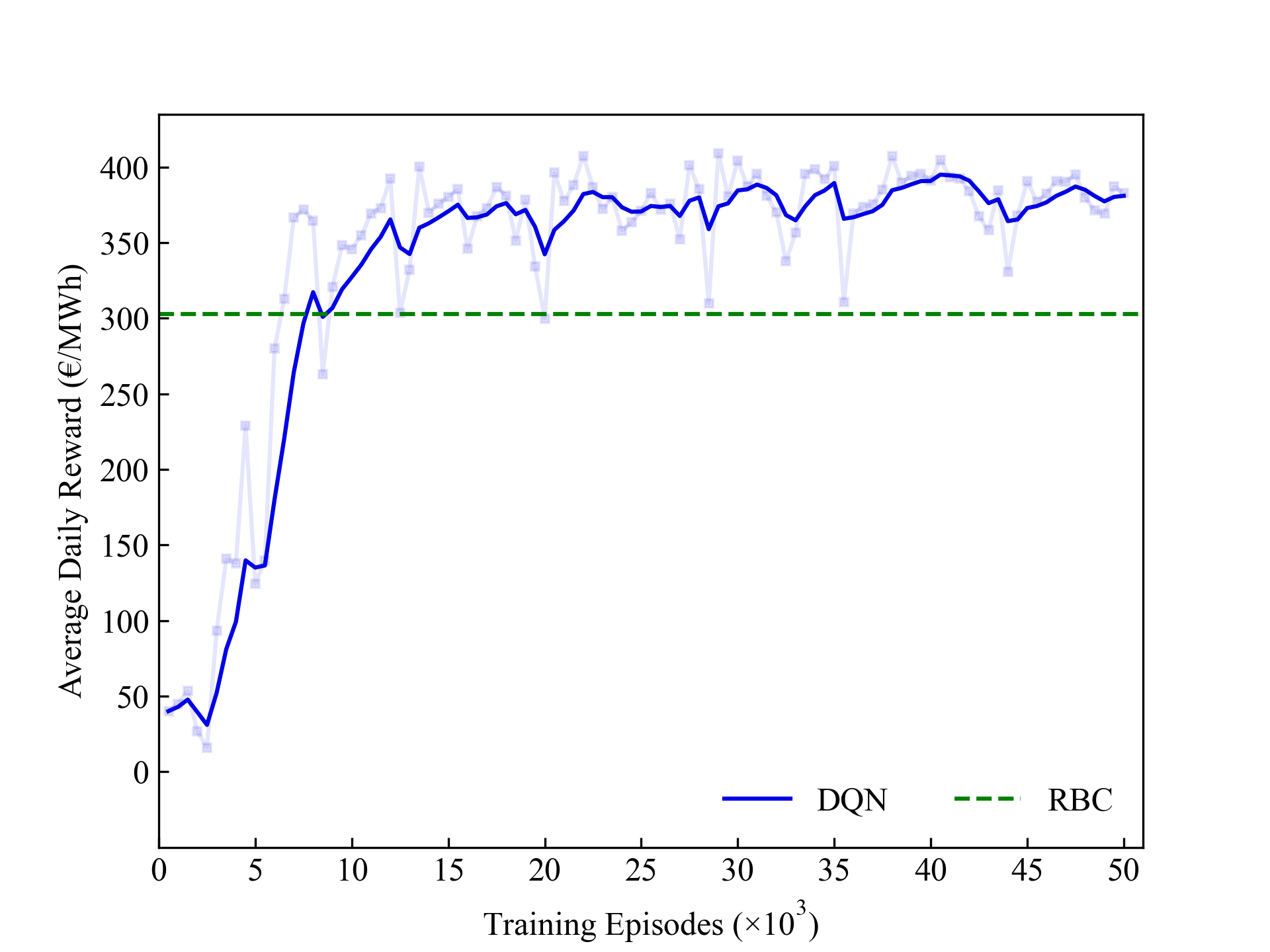}
        \caption{DQN Agent}
        \label{fig:learning process DQN}
    \end{subfigure}
    \hfill
    \begin{subfigure}{0.49\textwidth}
        \includegraphics[scale=0.5]{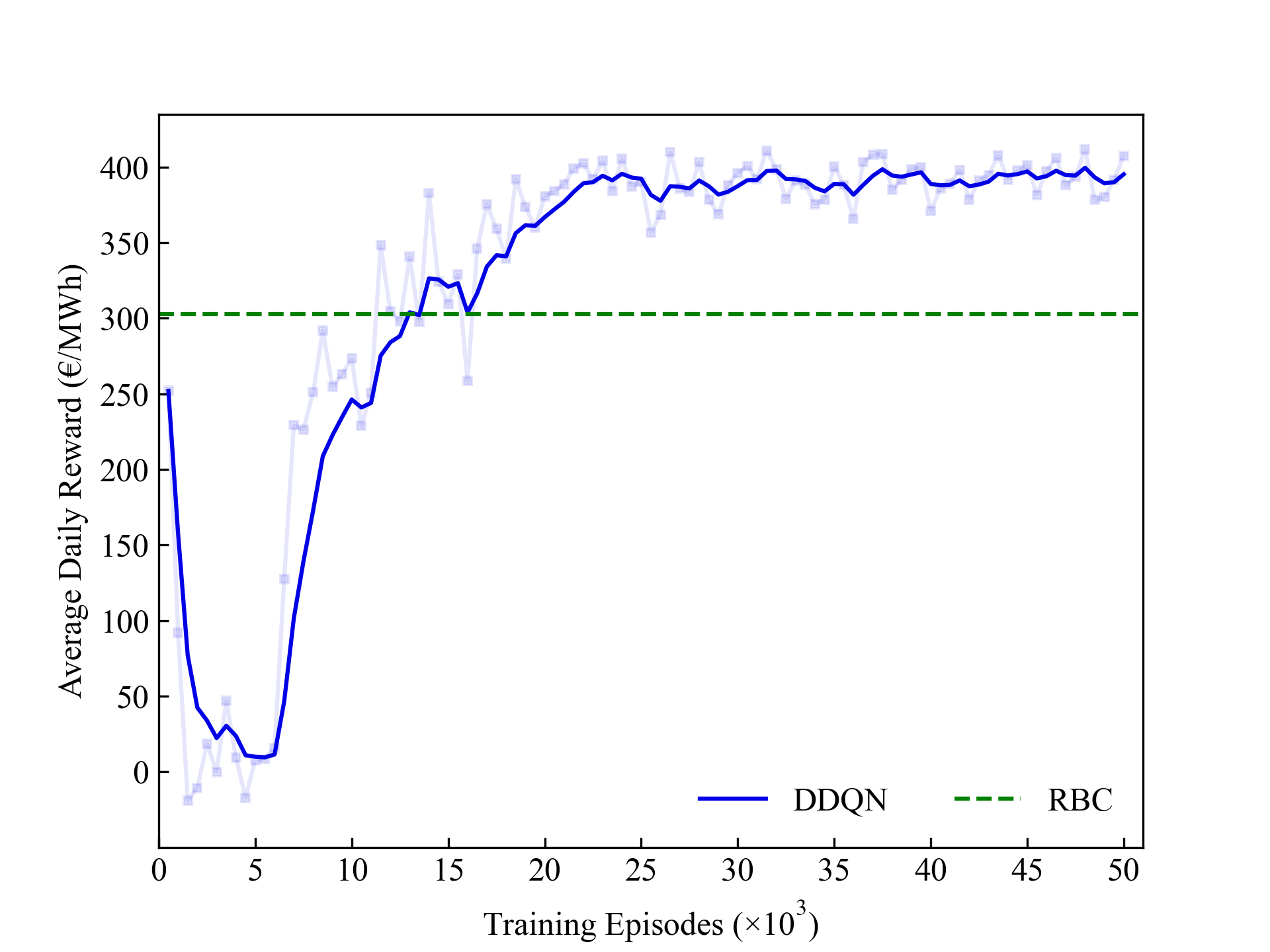}
        \caption{DDQN Agent}
        \label{fig:learning process DDQN}
    \end{subfigure}
    \caption{\myadd{The learning process of (a)~DQN and (b)~DDQN. The solid and faded lines show smoothed and actual learning curves, respectively.}}
    \label{fig:learning process}
\end{figure}

%--------------------------------
\myadd{The learning process of the RL methods is illustrated in \cref{fig:learning process}, %$. \Cref{fig:learning process} 
which shows the performance of the RBC and two RL methods on the validation set during the training.} \Cref{tab:simu profits} shows the performance of the \myadd{RBC and} trained RL methods on the test set. The DDQN method increases the average daily profit by \myadd{32.2\% and} 9.2\% compared to the \myadd{RBC and} DQN method\myadd{s, respectively}. The reason behind this is that the DDQN method estimates the probability distribution of returns instead of the expectation of returns. In this way, distributional RL methods mitigate instability in the Bellman optimality operator. The proposed post-processing step improves the performance by 3.2\% compared to that of the original DDQN model.

\begin{table}[h]
    \centering
    \caption{\myadd{Evaluation of RBC and RL methods on the test set.}}
    \begin{tabular}{ccccc}
        \toprule
        Method & \myadd{RBC} & DQN & DDQN & \begin{tabular}{@{}c@{}} DDQN after \\ policy correction \end{tabular} \\
        \midrule
        \begin{tabular}{@{}c@{}} Profit \\ (\texteuro/ day/ MWh)\end{tabular} & \myadd{341.1} & 413.1 & 450.9 & \textbf{465.2} \\
        \bottomrule
    \end{tabular}
    \label{tab:simu profits}
\end{table}

\begin{figure*}[h]
    \centering
    \begin{subfigure}{0.49\textwidth}
        \includegraphics[width=\linewidth]{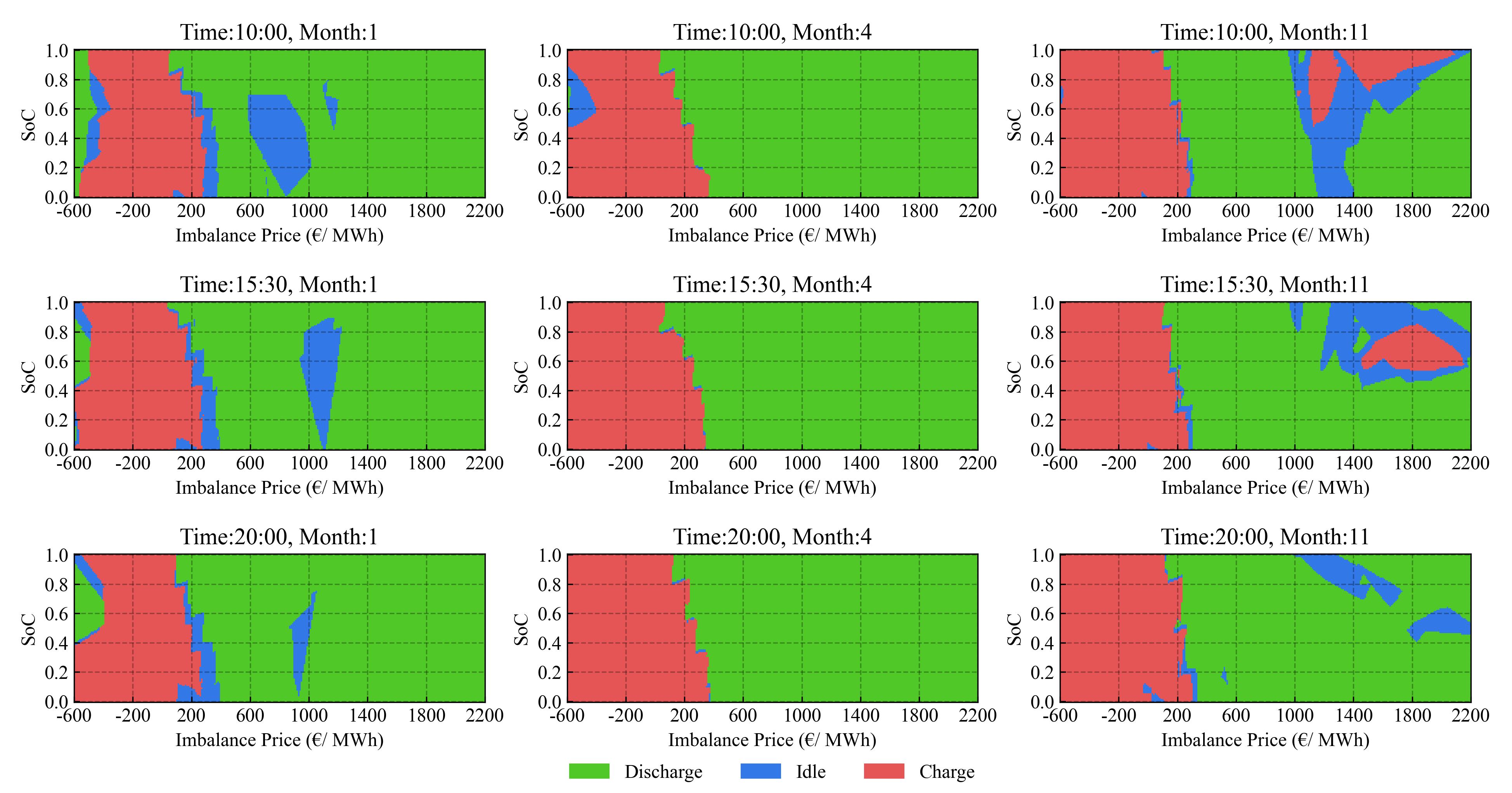}
        \caption{before}
        \label{fig:original policy}
    \end{subfigure}
    \hfill
    \begin{subfigure}{0.49\textwidth}
        \includegraphics[width=\linewidth]{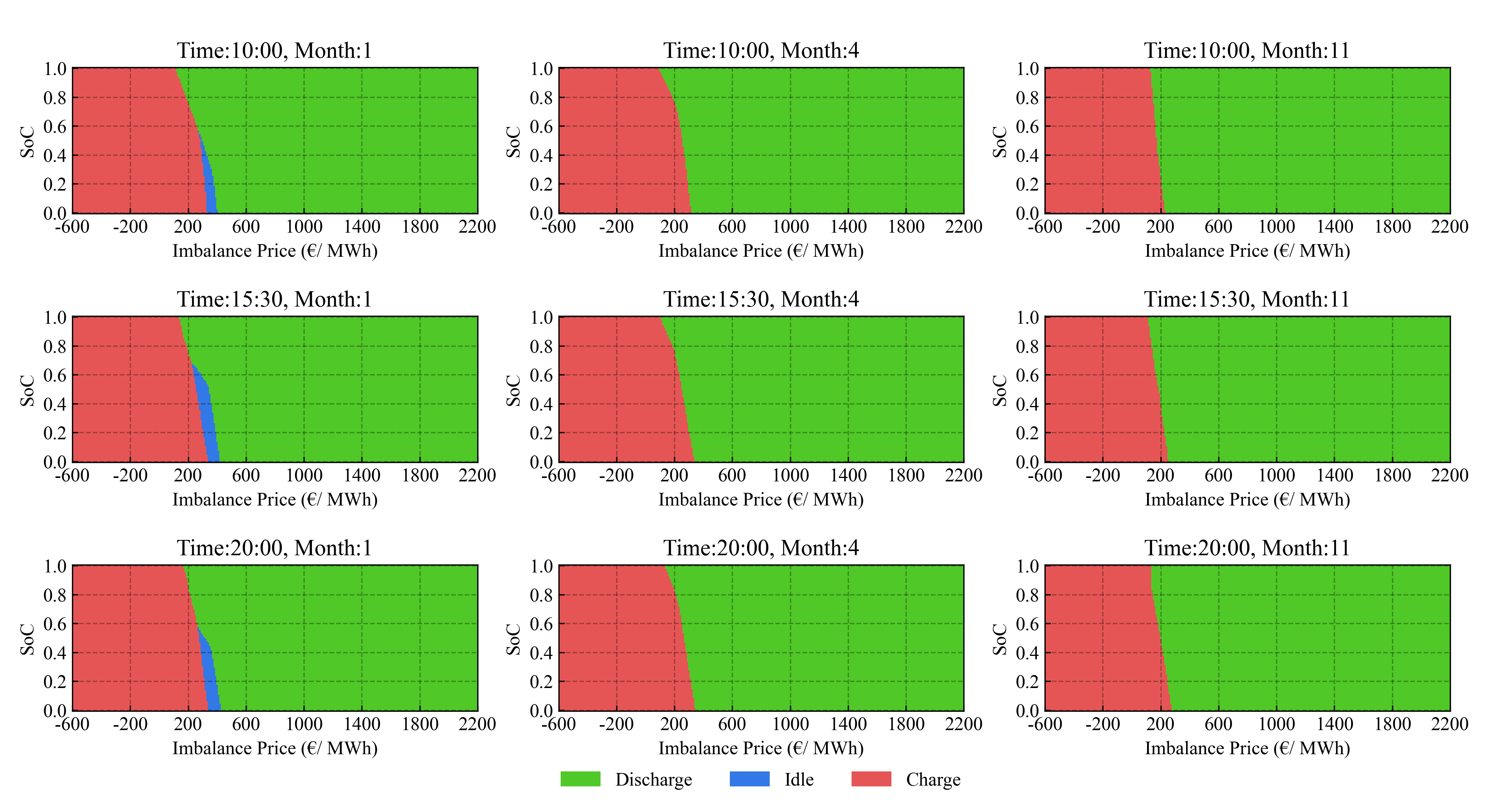}
        \caption{after}
        \label{fig:policy after correction}
    \end{subfigure}
    \caption{The learned policy (a) before and (b) after applying the policy correction step.}
    \label{fig:policies}
\end{figure*}

\begin{figure}[h]
    \centering
    \includegraphics[width=0.49\linewidth]{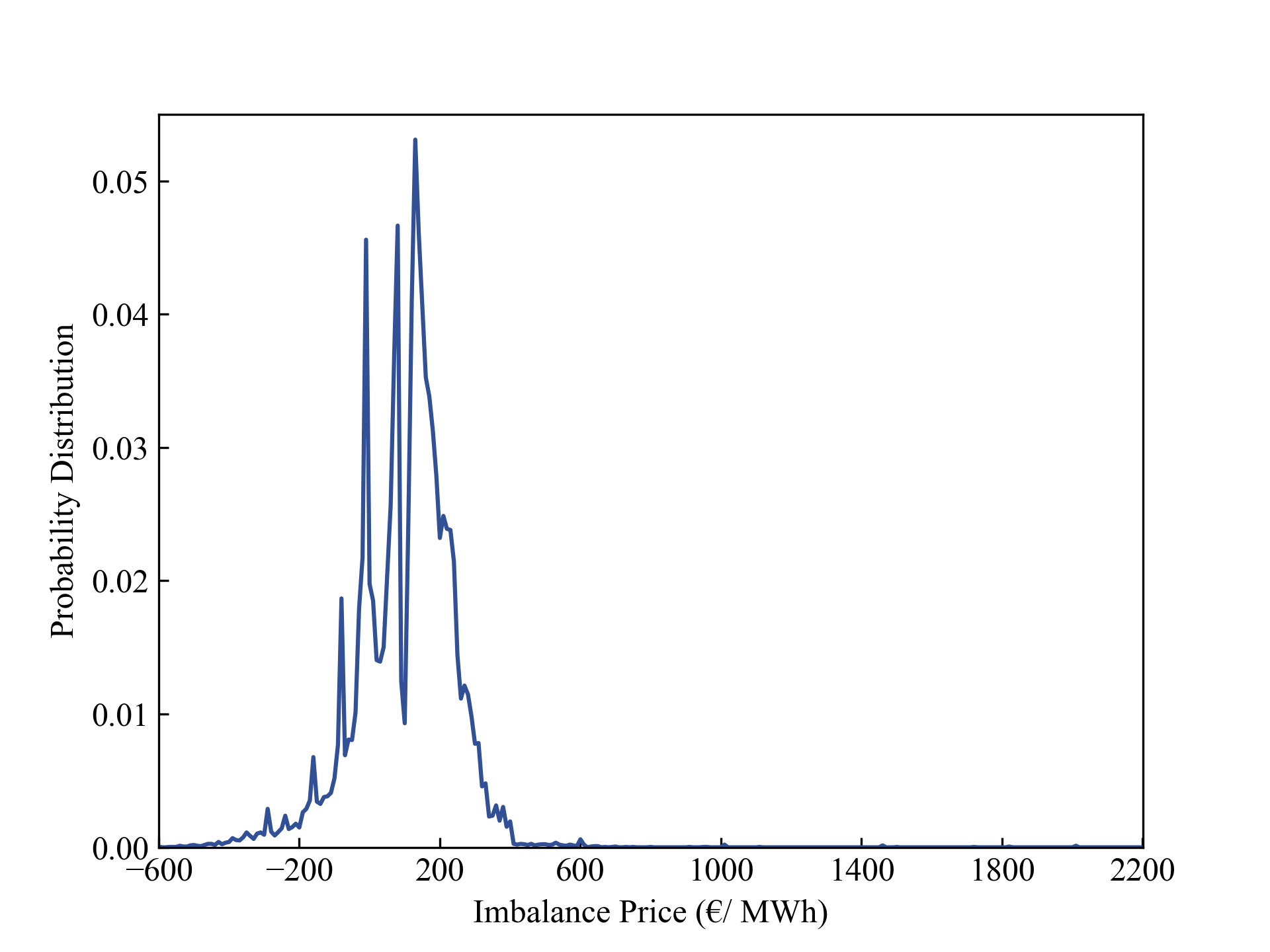}
    \caption{The probability distribution of the Belgian imbalance price in 2023.}
    \label{fig:imbalance probability distribution}
\end{figure}

\Cref{fig:policies} illustrates how the policy correction step improves the performance by showing the DDQN policy heatmaps before and after applying the policy correction step. As the most determinative features for the DDQN agent are \myadd{indicative} imbalance prices and SoC, we plot the learned policy with respect to these two input features for various times and months. It can be observed that the learned policy in some areas does not align with human intuition. For instance, there is a significant idle/ charge area in the policy for the 11th month at 10:00 when the price exceeds 1000 \texteuro /MWh, or there is a discharge region in the policy for the 1st month at 20:00 when the price is lower than $-$400 \texteuro /MWh. The reason for these imperfection areas can be found in the probability distribution of the price, shown in \cref{fig:imbalance probability distribution}. Prices below $-$200 \texteuro /MWh rarely occur (with a probability of 3\%), while prices above \num{1000} \texteuro /MWh are even rarer (with a probability of 0.04\%). Thus, the Q function $\mu_\theta$ overestimates Q values for out-of-distribution (OOD) actions in these rarely seen (or even unseen) states. As a result of this overestimation due to the max operator in the Bellman equation (\cref{9}), the agent takes unexpected actions in these rarely seen states. As \cref{fig:policy after correction} shows, after our post-processing step, all imperfections in the policy are removed. The distilled (corrected) policy successfully mimics the main decision thresholds of the teacher model, while replacing all OOD actions with correct actions. \myadd{As a result of replacing these OOD actions with correct actions, the student agent achieves higher performance than the original teacher model, as indicated in \cref{tab:simu profits}. Furthermore,} applying the policy correction step makes the final policy interpretable with clear decision boundaries.

\begin{figure}[h]
    \centering
    \includegraphics[width=0.49\linewidth]{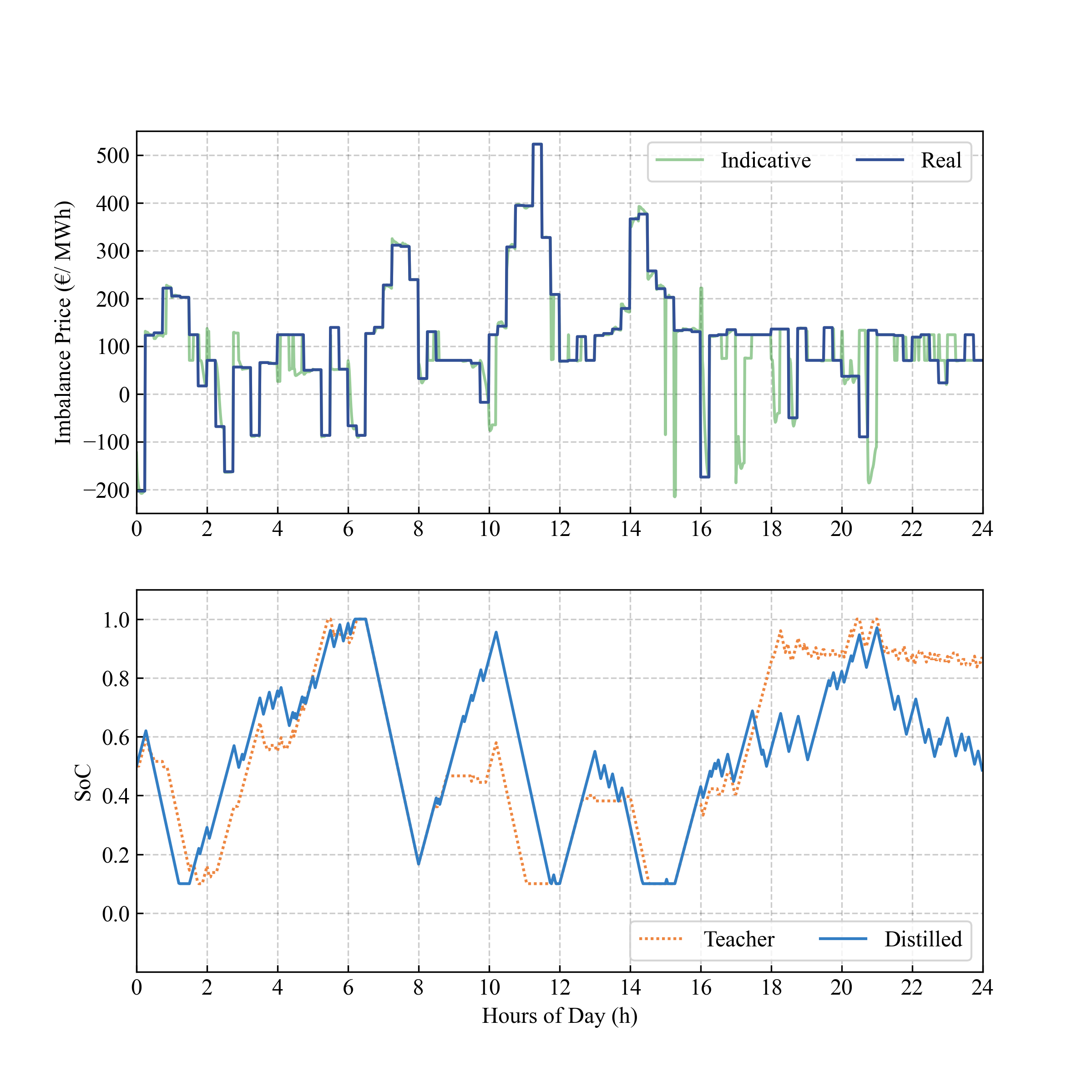}
    \caption{\myadd{The performance of the proposed controller on September 29, 2023.}}
    \label{fig:sample policy}
\end{figure}

To compare the performance of the agent before and after the post-processing step, the student and teacher models are tested using data from September 29, 2023. As \cref{fig:sample policy} shows, both models can effectively react to nearly all fluctuations in the price: they appropriately respond to two major peaks from 6:30 to 8:00 and from 10:30 to 12:00 by discharging the battery, and react to one major valley from 2:00 to 4:00 by charging the battery. However, the teacher model takes wrong actions between 8:30 and 10:00: the teacher agent decides to charge the battery between 8:15 and 8:30 when the price is approximately 130 \texteuro /MWh, while it does nothing from 8:30 to 10:00 when the price mostly hovers around 70 \texteuro /MWh and sometimes even dips to $-$20 \texteuro /MWh. On the other hand, the corrected model continuously charges the battery from 8:00 to 10:00, which ultimately results in a higher profit since the battery has more energy to discharge in the following hours where prices are higher (from 10:00 to 12:00).

%--------------------------------
\subsection{Experimental Results}
\label{sec:Experimental Results}
\begin{figure}[h]
    \centering
    \includegraphics[width=0.49\linewidth]{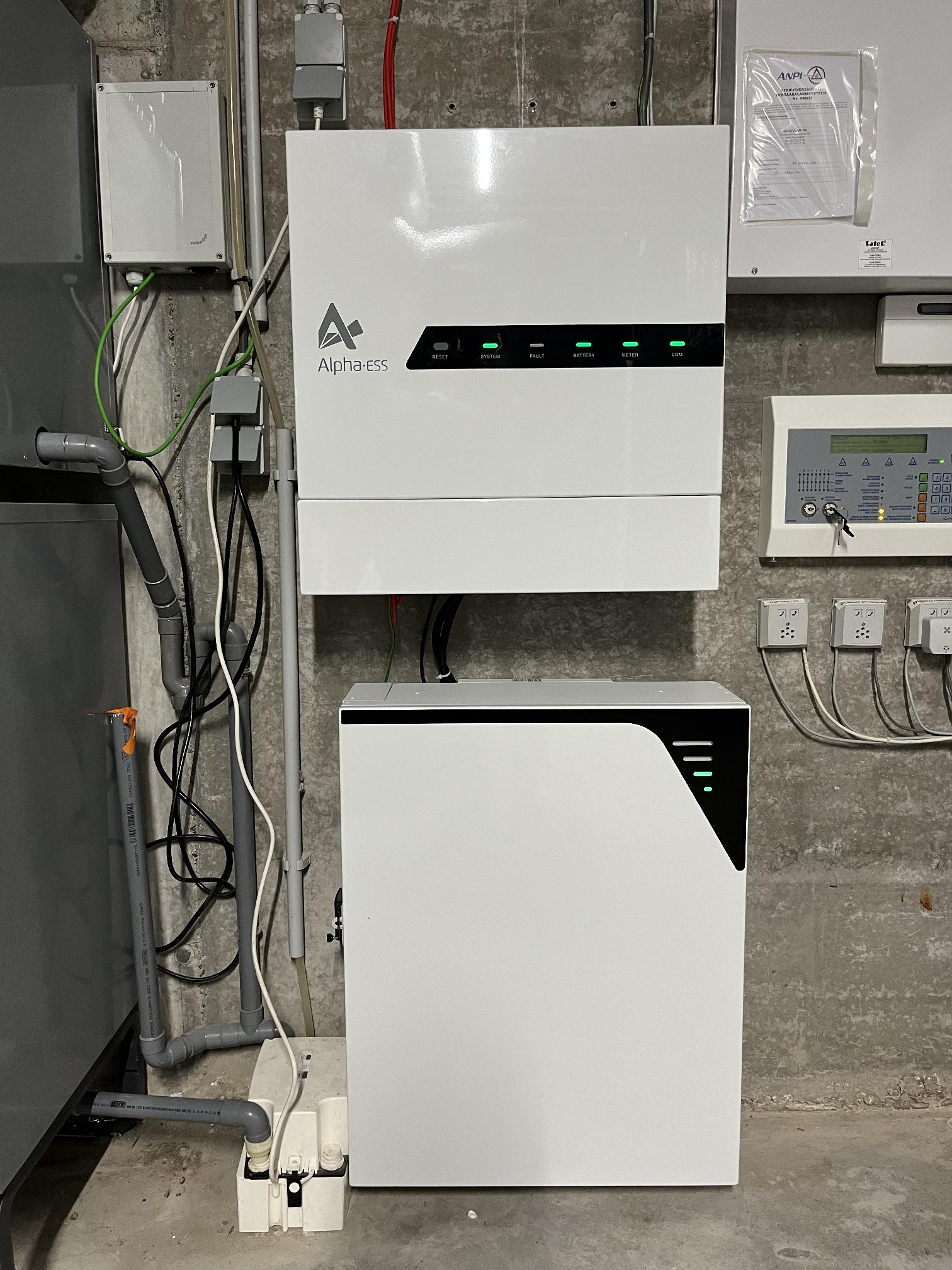}
    \caption{\textanon{HomeLab}{<Anonymous Experimental Facility>} battery used for testing the proposed control framework in real world}
    \label{fig:Homelab battery}
\end{figure}

%--------------------------------
We implement our proposed control framework on a real battery to better demonstrate its capabilities in a real-world setting. We use a residential AlphaESS battery with 4kW/ 8kWh, as shown in \cref{fig:Homelab battery}, which is installed in the \textanon{imec/ Ghent University HomeLab}{<Anonymous Experimental Facility>}. \textanon{HomeLab}{<Anonymous Experimental Facility>} is a real house offering a unique residential test environment for IoT services and smart home services.\footnote{\linkanon{https://homelab.ilabt.imec.be/}{https://homelab.ilabt.imec.be/}} For implementation, an API is available for the \textanon{HomeLab}{<Anonymous Experimental Facility>} which contains functionalities for the battery and other \textanon{HomeLabGym}{<Anonymous Experimental Facility>} devices~\cite{van2024homelabgym}.
This \textanon{HomeLab}{<Anonymous Experimental Facility>} API will forward the command to a battery-specific API. This codebase is running on a Raspberry Pi inside the \textanon{HomeLab}{<Anonymous Experimental Facility>}. The battery-specific codebase includes the translation needed from Python to the Modbus protocol in order to protocol to read and write appropriate registers on the battery energy management system (e.g., battery power) to the corresponding registers on the battery. The DDQN model, trained on the imbalance prices of 2023, after applying the post-processing step is used to control the \textanon{HomeLab}{<Anonymous Experimental Facility>} battery with a granularity of 2 minutes. We control the battery from January 21, 2024 to January 28, 2024 with Elia's real-time imbalance prices.

\begin{figure}[h]
    \centering
    \includegraphics[width=0.49\linewidth]{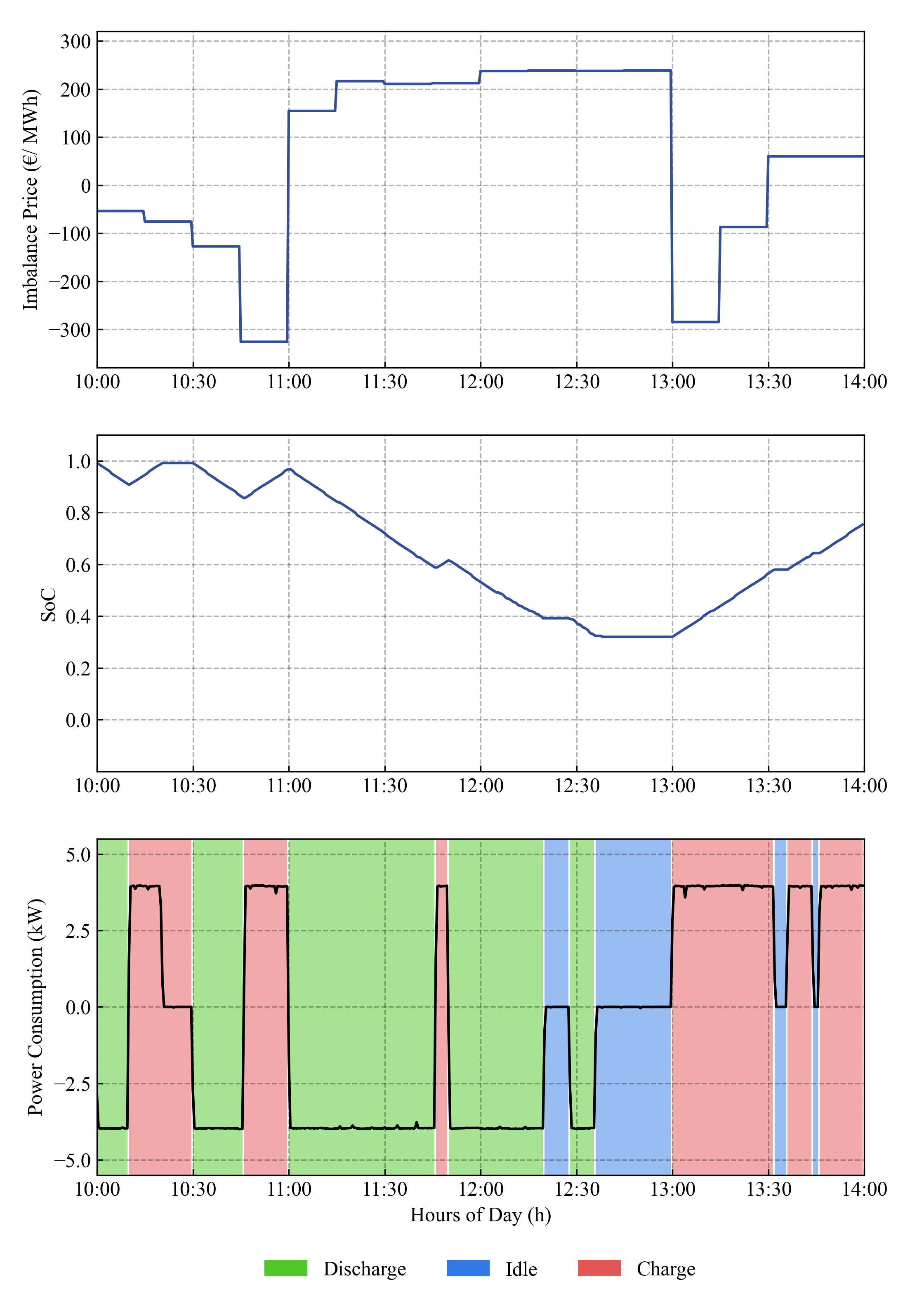}
    \caption{The experimental results on January 28, 2024 from 10:00 to 14:00.}
    \label{fig:experimental study}
\end{figure}

\Cref{tab:experimental profits} lists the average daily profit during the trial period. The revenue for the real-world implementation is 6.7\% lower than that for the simulation implementation (in the simulation implementation, we use the same price data as in the real-world implementation). There are three main reasons for this drop in the revenue: first, it takes 3s to take an action (includes doing calculations and fetch price data from Elia), 2s to send the taken action to the battery (communication delay), and finally, the battery needs an average of 5s to change it. Hence, it takes an average of 10s to select and execute the appropriate action, which accounts for nearly 8\% of the 2-minute time step. Second, from January 22 at 22:00 to January 23 at 7:00, Elia faced technical issues that caused a delay in publishing imbalance price data. This led to our controller not receiving the necessary values to take appropriate actions. Third, sometimes excessive frequent changes between charging and discharging can cause the temperature of the battery to rise, resulting in decreased battery efficiency. \Cref{fig:experimental study} demonstrates a snapshot of the experimental results on January 28 between 10:00 and 14:00. The bottom row figure displays the actual power consumed by the \textanon{HomeLab}{<Anonymous Experimental Facility>} battery, with colors indicating the actions sent to the battery. This experimental study successfully demonstrated the deployability of our framework. 

\begin{table}[h]
    \centering
    \caption{Evaluation of the proposed control framework on the real battery.}
    \begin{tabular}{cccc}
        \toprule
        Method & \begin{tabular}{@{}c@{}} DDQN after \\ policy correction \\ (simulation) \end{tabular} & \begin{tabular}{@{}c@{}} DDQN after \\ policy correction \\ (real world) \end{tabular}\\
        \midrule
        \begin{tabular}{@{}c@{}} Profit \\ (\texteuro/ day)\end{tabular} & 6 & 5.6 \\
        \bottomrule
    \end{tabular}
    \label{tab:experimental profits}
\end{table}

%===================================
\section{Conclusion}
\label{sec:Conclusion}
%===================================
In this paper, we proposed a new RL-based control framework for batteries to perform energy arbitrage in the imbalance settlement mechanism. In our proposed control framework, we first train the agent to maximize the arbitrage revenue. Then, the post-processing step corrects (constrains) the learned policy during a knowledge distillation process based on defined human-intuitive properties. We extended the standard knowledge distillation process by incorporating an optimization layer into the student model and modifying the standard distillation loss to remove the (slow) differentiable layer at inference. The performance of the proposed control framework was evaluated through both simulation and experimental results using the Belgian imbalance price of 2023. The results, using DDQN as the RL algorithm, demonstrated that adding the post-processing step outperforms \myadd{all of} the \myadd{RBC,} vanilla DQN\myadd{, and vanilla} DDQN methods. The DDQN method with the policy correction step could improve the average daily profit by \myadd{36.4\% and} 12.6\% compared to the \myadd{RBC and} DQN method\myadd{s, respectively}. This improvement stems from two factors: 
\begin{enumerate*}[(i),nosep,noitemsep]
    \item the \textit{distributional} perspective diminishes instability in the Bellman optimality operator by learning the full probability distribution of returns rather than a single value expectation of returns;
    \item the \textit{post-processing} step deals with Q-value overestimation by replacing OOD actions with correct (human-intuitive) actions.
\end{enumerate*}
During the post-processing step, the student model effectively distilled the knowledge from the pretrained teacher model regarding the main decision boundaries, while correcting the policy based on the defined human-intuitive properties. We deploy our proposed control framework in a real-world experimental setup, i.e., on the \textanon{HomeLab}{<Anonymous Experimental Facility>} battery to investigate its performance in a real-world environment. The experimental performance is 6.7\% lower than the simulated environment, due to delays in action calculation and execution, missing data in real-time imbalance prices, and a decrease in battery efficiency as a result of temperature rise. Note that although we used DDQN in this paper, the proposed control framework is applicable to all RL methods.  

In future research, we will focus on developing an online knowledge distillation process wherein both teacher and student models are trained end-to-end during the RL training loop. Another direction for future work is to consider adding more constraints to the proposed control policy, such as a (daily) cycle constraint for the battery.

\printbibliography
\end{document}